\documentclass[10pt,journal,compsoc]{IEEEtran}

\ifCLASSOPTIONcompsoc
  \usepackage[nocompress]{cite}
\else
  \usepackage{cite}
\fi

\usepackage{amsmath,amsfonts}
\usepackage{array}
\usepackage{textcomp}
\usepackage{stfloats}
\usepackage{url}
\usepackage{verbatim}
\usepackage{graphicx}
\usepackage{cite}
\usepackage[utf8]{inputenc}
\usepackage{dsfont}
\usepackage{amsfonts}
\usepackage{amsmath}
\usepackage{amssymb}
\usepackage[ruled]{algorithm2e}
\usepackage{soul}
\usepackage{bm}
\usepackage{color, xcolor}
\soulregister\cite7
\soulregister\citep7
\soulregister\ref7
\usepackage{booktabs}

\usepackage{changepage}
\usepackage{svg}
\usepackage{multirow}
\usepackage{graphicx}
\usepackage{caption}
\usepackage{subcaption} 
\usepackage{mathrsfs}
\usepackage{threeparttable}

\usepackage{tabularx}
\usepackage{pgfplots}
\pgfplotsset{compat=newest}

\usepackage{tikz}
\usetikzlibrary{positioning, calc, shapes.geometric, shapes.multipart, backgrounds,
shapes, arrows.meta, arrows,
decorations.markings, external, trees}
\tikzstyle{Arrow} = [
thick,
decoration={
markings,
mark=at position 1 with {
\arrow[thick]{latex}
}
},
shorten >= 3pt, preaction = {decorate}
]
\newcommand{\mG}{\mathcal{G}}
\newcommand{\mI}{\mathcal{I}}

\newcounter{Counter1}
\newcommand{\Rone}{\roman{Counter1}}

\newcounter{Counter2}
\newcommand{\Rtwo}{\roman{Counter2}}

\newcounter{Counter3}
\newcommand{\Rthree}{\roman{Counter3}}

\newcounter{Counter4}
\newcommand{\bH}{\mathbf{H}}
\newcommand{\bQ}{\mathbf{Q}}
\newcommand{\bK}{\mathbf{K}}
\newcommand{\bV}{\mathbf{V}}
\newcommand{\bA}{\mathbf{A}}
\newcommand{\bP}{\mathbf{P}}
\newcommand{\bF}{\mathbf{F}}
\newcommand{\bT}{\mathbf{T}}

\newcommand{\bI}{\mathbf{I}}

\newtheorem{mydef}{\textbf{Definition}}

\begin{document}

\title{Fusing LLMs and KGs for Formal Causal Reasoning behind Financial Risk Contagion}

\author{Guanyuan Yu,
        Xv Wang,
	    *Qing Li~\IEEEmembership{Member,~IEEE},
	    Yu Zhao~\IEEEmembership{Member,~IEEE}
      

	\IEEEcompsocitemizethanks{\IEEEcompsocthanksitem All authors
	are with Southwestern University of Finance and Economics, Chengdu, China.
	\IEEEcompsocthanksitem Corresponding author Qing Li (E-mail: liq\_t@swufe.edu.cn).
	
	\IEEEcompsocthanksitem This research is supported by the National Natural Science Foundation of China under Grant No. 62072379 and 62376227, Sichuan Science and Technology Program under Grant No. 2023NSFSC0032.
	
	}

}

\IEEEtitleabstractindextext{
\begin{abstract}
Financial risks trend to spread from one entity to another, ultimately leading to systemic risks. The key to preventing such risks lies in understanding the causal chains behind risk contagion. Despite this, prevailing approaches primarily emphasize identifying risks, overlooking the underlying causal analysis of risk.
To address such an issue, we propose a \underline{\textbf{R}}isk \underline{\textbf{C}}ontagion \underline{\textbf{C}}ausal \underline{\textbf{R}}easoning model called \textbf{RC\textsuperscript{2}R}, which uses the logical reasoning capabilities of large language models (LLMs) to dissect the causal mechanisms of risk contagion grounded in the factual and expert knowledge embedded within financial knowledge graphs (KGs).
At the data level, we utilize financial KGs to construct causal instructions, empowering LLMs to perform formal causal reasoning on risk propagation and tackle the "causal parrot" problem of LLMs.
In terms of model architecture, we integrate a fusion module that aligns tokens and nodes across various granularities via multi-scale contrastive learning, followed by the amalgamation of textual and graph-structured data through soft prompt with cross multi-head attention mechanisms.
To quantify risk contagion, we introduce a risk pathway inference module for calculating
risk scores for each node in the graph.
Finally, we visualize the risk contagion pathways and their intensities using  Sankey diagrams, providing detailed causal explanations. 
Comprehensive experiments on financial KGs and supply chain datasets demonstrate that 
our model outperforms several state-of-the-art models in prediction performance and out-of-distribution (OOD) generalization capabilities.
We will make our dataset and code publicly accessible to encourage further research and development in this field.

\end{abstract}

}

\maketitle
\IEEEdisplaynontitleabstractindextext
\IEEEpeerreviewmaketitle

\begin{figure}[htbp]
	\centering
    \includegraphics[width=0.55\textwidth]{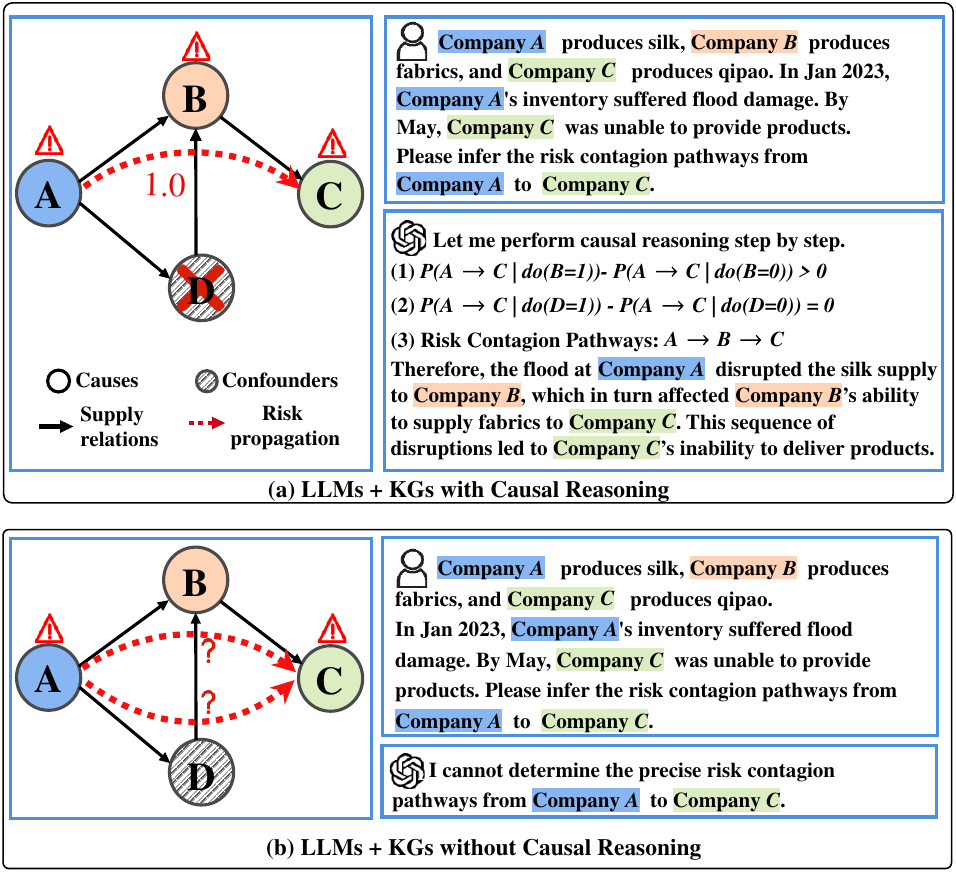}
	\caption{A Case of Risk Causal Reasoning. (a) It utilizes formal causal inference to eliminate confounders as well as accurately trace and quantify the pathways of risk contagion, whereas (b) fails to accomplish this.}
	\label{fig:supply-chain-risk}
\end{figure}

\IEEEraisesectionheading{\section{Introduction}}
\IEEEPARstart{F}inancial risk contagion, the phenomenon where the risk from one financial entity rapidly spreads to others, can escalate into systemic risks if not properly controlled~\cite{gai2010contagion,elliott2014financial,eisenberg2001systemic,billio2012econometric}.
The essence of financial risk control lies in a profound understanding of the causal mechanisms behind risk contagion.
As shown in Fig.~\ref{fig:supply-chain-risk}(a), through formal causal inference, we can ensure that the causal chain of risk contagion is $A\rightarrow B \rightarrow C$.
Consequently, 
we can design effective strategies to block the chain of risk contagion.
However, 
the majority of existing studies concentrate on the development of risk recognition models based on machine learning~\cite{gepp2015predicting, malekipirbazari2015risk,danenas2015selection,chang2016establishing,jiang2018stationary,arora2020bolasso} and deep learning~\cite{fischer2018deep,hosaka2019bankruptcy,wang2019semi,li2020multimodal,xu2021towards,ouyang2021systemic,cheng2022regulating,cheng2022financial}.
They neglect the intricate causal mechanisms behind financial risk contagion, hindering the implementation of risk prevention and control strategies.

To address such issues, this study proposes a novel
\underline{\textbf{R}}isk \underline{\textbf{C}}ontagion \underline{\textbf{C}}ausal \underline{\textbf{R}}easoning model called \textbf{RC\textsuperscript{2}R},
which uses the logical reasoning capabilities of large language models (LLMs)~\cite{zhang2023understanding,webb2023emergent, yang2024give, chen2024learning}  to analyze the causal mechanisms behind risk contagion, based on factual and professional knowledge in the financial knowledge graphs (KGs)~\cite{loster2018challenges, pan2024unifying}.
Specifically,
as shown in Fig.~\ref{fig:supply-chain-risk}(a), we first perform random interventions on the financial KGs, and then guide the LLMs to conduct formal causal reasoning to identify the risk contagion paths. 
Based on information retrieved from KGs, 
LLMs generate explanations for queries while quantifying the intensity of risk propagation.
For instance, to determine if node $B$ is part of the financial risk contagion path, we verify the change in contagion probability before and after intervening at node $B$. A greater change suggests a higher likelihood of node $B$ being on the contagion path, while a smaller change indicates its lesser involvement.

Notably, the data foundation for LLMs' inference is natural language, whereas financial KGs are structured as graphs, leading to significant disparities in data modality. Here, we seamlessly integrate these two modalities from two perspectives.
\textbf{(\Rone) At the data level}, to guide LLMs in performing formal causal reasoning, we develop causal instructions. Such instructions include prompts, queries, and explanations, all based on financial KGs. Both KGs and instructions are then inputted into GNNs and LLMs,
generating high-level representations.
\textbf{(\Rtwo) At the architectural level}, we introduce a fusion module that employs a multi-scale contrastive loss function to explicitly coordinate the alignments between textual data and financial KGs. This module then leverages soft prompt with cross multi-head attention mechanisms to facilitate interaction from LLMs to KGs. Following this, a risk pathway inference module is implemented to identify and analyze predominant risk propagation pathways effectively.

During the inference stage, LLMs
trace the pathways of financial risk contagion
on KGs according to queries,
generating explanations.
Then, we can utilize Sankey diagrams to visualize the direction and intensity of risk contagion.
In essence, we employ LLMs as a causal reasoning engine to explore risk contagion paths within financial KGs.
In summary, our study makes the following three unique contributions.

    $\bullet$~ To the best of our knowledge, this study represents the first effort to activate the formal causal reasoning capabilities of LLMs while utilizing the factual and specialized knowledge embedded in financial KGs. This approach provides profound insights into the causal mechanisms of risk contagion and opens up innovative methodologies and perspectives for financial risk analysis.
    
    $\bullet$~
    We introduce multi-scale contrastive learning to align tokens and nodes, and employ soft prompt with cross multi-head attention to seamlessly integrate text and graph information.
    In the risk pathway inference module, we introduce intervention mechanisms to precisely estimate the impact of each variable on risk contagion.
    
    $\bullet$~Our model demonstrates excellent performance and successfully visualizes the pathways of risk contagion. We have made our datasets and code publicly available for further research and development\footnote{Please refer to https://github.com/wang1595243339/RC2R.}.
    To our knowledge, we are the first to open-source comprehensive datasets for causal analysis of financial risk contagion, including causal corpora and financial KGs.

\section{Related Work}

\subsection{Financial Risk Detection \& Analysis}
Current studies usually leverage deep learning models for risk detection and analysis.
For example,
many unified models based on the convolutional neural networks (CNNs) and long short-term memory (LSTM)
are introduced to process financial data, providing insights into market fluctuations~\cite{lu2020cnn,chen2021constructing,cavalli2021cnn,wang2021stock}.
The graph neural network (GNN) family
(e.g., 
Risk-Rate~\cite{cheng2020contagious}, 
Know-GNN~\cite{rao2021know}, MAGNN~\cite{cheng2022financial},  DGA-GNN~\cite{duan2024dga})
is utilized to analyze trading data within financial networks, identifying unusual transactions and financial risks.
Besides, 
Transformer and its variants (e.g., 
Html~\cite{yang2020html},
Numhtml~\cite{yang2022numhtml},
FinBERT~\cite{huang2023finbert},
GPT-3~\cite{leippold2023sentiment}) are applied to analyze financial risks.
However, these models often overlook the complex causal mechanisms behind financial risk contagion, hindering effective risk prevention and control strategies. To address this issue, we propose integrating LLMs and KGs to investigate the inherent causal mechanisms of risk contagion.

\subsection{Causal Inference}
Classical studies have utilized deep learning models to identify causal structures and separate causal variables~\cite{brehmer2022weakly,wang2022causal}. For instance, deep structural causal models (DSCMs) employ normalizing flows and variational inference for tractable inference of exogenous noise variables~\cite{pawlowski2020deep}. Techniques like DIR~\cite{wu2021discovering} and V-REx~\cite{krueger2021out} are used to differentiate causal from non-causal features in input data.

Nowadays, more and more studies concentrate on using LLMs to discover causal relationships and estimate causal effects~\cite{naik2023applying,kiciman2023causal,feder2024causal,liu2024large}.
For example, 
GPT family (e.g., GPT-3~\cite{long2023can}, ChatGPT~\cite{tu2023causal}) is used to answer causal discovery questions and look for causal directions.
DISCO makes the estimation of
causal effects by generating counterfactual data~\cite{chen2022disco}. CInA performs self-supervised causal learning and facilitates zero-shot causal inference~\cite{zhang2023towards}.

As an advancement of the above studies, we are the first to explore causal discovery and estimation in financial risk contagion by integrating LLMs with KGs. Our approach provides deep insights into financial risk contagion and aids in the development of risk prevention and control strategies.

\subsection{Fusion of LLMs and KGs}
LLMs have been shown to exhibit hallucinations~\cite{gunjal2024detecting}. To address this issue, some studies have focused on integrating the specialized and factual knowledge of KGs into the inference process of LLMs~\cite{pan2024unifying,guan2024mitigating}. Broadly, there are two popular approaches to this integration.

The first approach treats LLMs as intermediary agents that interact with KGs~\cite{zeng2023agenttuning}. For instance, StructGPT~\cite{jiang2023structgpt} introduces special interfaces that enable LLMs to access and reason with information from KGs. Similarly, Think-on-graph~\cite{sun2023think} identifies reasoning paths to elucidate how LLMs infer and generate answers.

The second approach augments LLMs with KG knowledge during the training or fine-tuning phases. This typically involves using separate encoders for text and graph-based data, as seen in models like KagNet~\cite{lin2019kagnet} and QA-GNN~\cite{yasunaga2021qa}. JointLK~\cite{sun2021jointlk} further refines this method by scoring reasoning results at each step through a bi-directional attention mechanism.

In our study, to mitigate hallucinations in LLMs during causal reasoning, we enhance the integration of LLMs and KGs by employing a multi-scale contrastive loss function and soft prompts with cross multi-head attention mechanisms.

\section{Our Approach}

\begin{figure*}[htbp]
	\centering
	\includegraphics[width=0.7\textwidth]{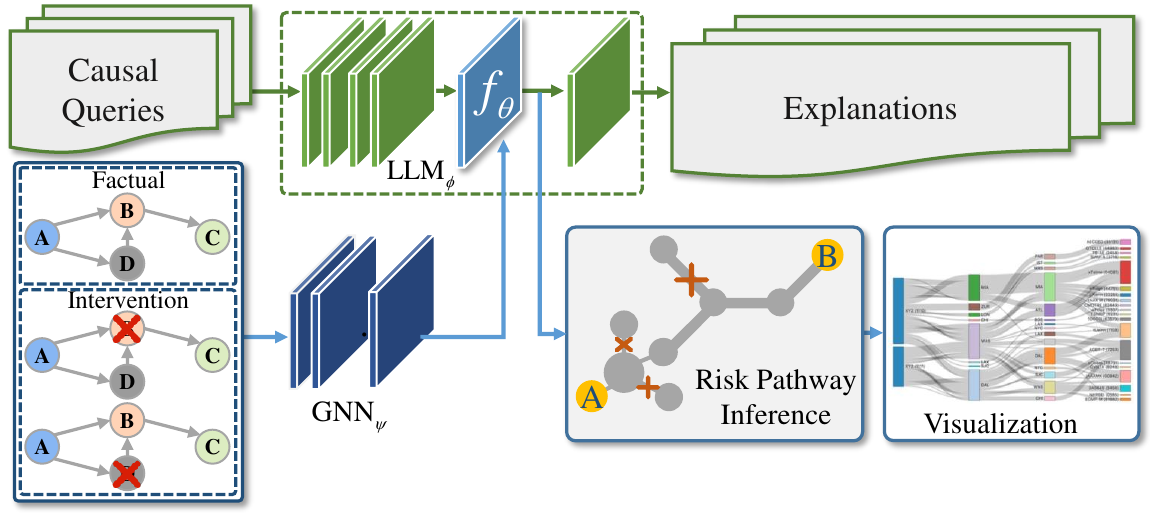}
	\caption{The Whole Framework of Our Proposed RC\textsuperscript{2}R}
	\label{fig:llm-kg-overview}
\end{figure*}

Fig.~\ref{fig:llm-kg-overview} provides an overview of RC\textsuperscript{2}R. Initially, casual queries and financial KGs are processed through a large language model ${\rm LLM}_\phi(\cdot)$ and a graph neural network ${\rm GNN}_\psi(\cdot)$, respectively, to produce high-level representations. These representations are then integrated in a latent space by a fusion module, denoted as $f_\theta(\cdot)$. Following this integration, ${\rm LLM}_\phi(\cdot)$ leverages such representations to generate responses and corresponding explanations. Additionally, a risk pathway inference module is employed to identify the predominant propagation pathways. To depict the direction and magnitude of risk contagion, Sankey diagrams are finally utilized.

Notably, the key to achieving the aforementioned objective lies in the application of LLMs to accurately identify the ground-truth causal relationships of risk propagation within financial KGs.
Nevertheless, 
LLMs can generate text that appears coherent and logically sound, effectively mimicking causal reasoning without truly grasping the underlying causal relationships. Such responses are based on pattern recognition from extensive training datasets rather than an actual understanding of real-world dynamics or independent logical reasoning,
that is, LLMs may be "causal parrots"~\cite{zevcevic2023causal,jin2024cladder}.
%

For the first time, we innovatively employ the following three steps to pioneer the activation of causal reasoning capabilities in LLMs.
(\Rone) We develop a formal causal diagram (Fig.~\ref{fig:dags}) to understand, analyze, and predict the causal relationships of risk contagion. (\Rtwo) In accordance with well-defined formal rules of causal inference, 
we formulate various causal queries and provide detailed explanations to guide LLMs in recognizing cause-and-effect relationships in risk contagion, as depicted in Table~\ref{tab:instructions}.  
(\Rthree) We seamlessly integrate LLMs with GNNs via our proposed fusion module.

\subsection{Formal Causal Reasoning for Risk Contagion}
\label{sec:formal causal reasoning}
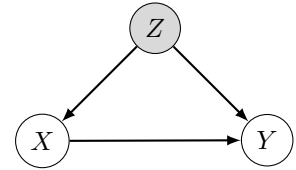
\begin{figure}[htbp]
    \centering
    
    \begin{tikzpicture}
    \node [circle, draw, fill=gray!30] (Z) {$Z$};
    \node [circle, draw, below left =of Z] (X) {$X$};
    \node [circle, draw, below right =of Z] (Y) {$Y$};

    \draw[Arrow, thick] (Z) -- (Y);
    \draw[Arrow, thick] (Z) -- (X);
    \draw[Arrow, thick] (X) -- (Y);

    \end{tikzpicture}

    \caption{Visualization of SCM via a Directed Acyclic Graph}
   \label{fig:dags}
\end{figure}

Leveraging the inference capabilities of LLMs to conduct an  in-depth analysis of risk propagation mechanisms in financial KGs primarily involves identifying causal variables within these graphs. Here, let $\mG=\{X, Z\}$ denote the KGs, where $X$ represents the causal components, i.e., factors influencing risk contagion, and $Z$ signifies the non-causal components, i.e., factors not affecting risk contagion. Meanwhile, $Y$ is the outcome variable that reflects the result of risk propagation.
Here, we utilize the Structural Causal Model (SCM)~\cite{pearl2016causal} to thoroughly explore the causal relations between these variables.
Fig.~\ref{fig:dags} illustrates two types
of relations:
(\Rone) $X\rightarrow Y$ means that the casual part $X$ is the endogenous parent responsible for determining $Y$.
Taking the risk contagion in Fig.~\ref{fig:supply-chain-risk}(a) as an example,
$\textit{Company A}\rightarrow \textit{Company B} \rightarrow \textit{Company C}$ is the casual part, which
 perfectly explains the mechanism of risk contagion. 
(\Rtwo) $X \leftarrow Z\rightarrow Y$ reflects that
$Z$ is a confounder between $X$ and $Y$, which opens a backdoor path.
To discern the ground-truth causal relationships of risk contagion, we need to block such a backdoor pathway.

Here, we attempt to integrate LLMs and financial KGs
to block the backdoor paths. Specifically,
\begin{equation*}
    \begin{split}
        Y&=f_Y(X), Y \perp \!\!\! \perp Z \,|\, X, \\
        s.t.~X&= f_\theta({\rm LLM}_\phi(\mathcal{I}),{\rm GNN}_\psi(\mG_\mathrm{set})).
    \end{split}
\end{equation*}
In the above equation, ${\rm LLM}_\phi(\cdot)$ represents a large language model equipped with a parameter set denoted by $\phi$. ${\rm GNN}_\psi(\cdot)$ refers to a graph neural network characterized by its parameter set $\psi$. $\mathcal{I}$ corresponds to the causal instruction, depicted in Table~\ref{tab:instructions}. 
$\mG_\mathrm{set}$ denotes the set of factual and intervention
graphs in Algorithm~\ref{algo:intervention}.
The function $f_\theta(\cdot, \cdot)$ is responsible for integrating the instruction information and graph information within a latent space.

\subsection{Casual Instruction \& Data Intervention}

\begin{table}[htbp]
\centering
\caption{Causal Instruction \& Data Invervention}
\begin{threeparttable}
\begin{tabular}{ccc}
\toprule

\multicolumn{3}{p{8.5cm}}{\textbf{Role:} You are a professional assistant specializing in the causal reasoning of financial risk contagion. Please use your knowledge of financial risk to answer the following questions.} \\

\multicolumn{3}{p{8.5cm}}{\textbf{Query:} 
{\color{blue}\textit{Company A}} produces silk, {\color{orange} \textit{Company B}} produces fabrics, and {\color{green}\textit{Company C}} produces qipao.
In Jan 2023, {\color{blue}\textit{Company A}}'s inventory suffered flood damage. By May, {\color{green}\textit{Company C}} was unable to provide products. 
Please infer the risk contagion pathways from {\color{blue}\textit{Company A}} to {\color{green}\textit{Company C}}.}
\\

\midrule
\textbf{Groups} &
\textbf{Graphs} &
\textbf{Contagion Probabilities}
\\

\midrule
Factual\tnote{*}
&
\raisebox{-0.5\height}{\includegraphics[width=0.8in]{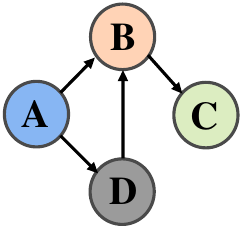}}
&
\begin{tabular}[c]{@{}c@{}}$P\left(A\rightarrow C | do(A=1)\right)=1$\\
$P\left(A\rightarrow C | do(B=1)\right)=1$\\
$P\left(A\rightarrow C | do(D=1)\right)=1$
\end{tabular}\\

\midrule

\multirow{12}{*}{Intervention}
&
\raisebox{-0.5\height}{\includegraphics[width=0.8in]{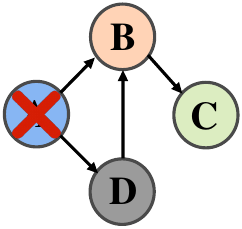}}
&
$P(A\rightarrow C|do(A=0))=0$

\\\cmidrule{2-3}

&
\raisebox{-0.5\height}{\includegraphics[width=0.8in]{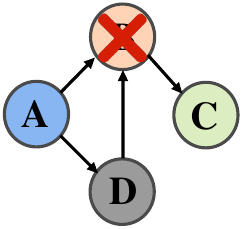}}
&
$P(A\rightarrow C | do(B=0))=0$

\\\cmidrule{2-3}
& 
\raisebox{-0.5\height}{\includegraphics[width=0.8in]{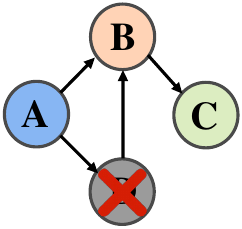}}
& $P(A\rightarrow C | do(D=0))=1$
\\
\midrule
\multicolumn{3}{p{8.5cm}}{\textbf{Explanation:} Let me perform formal causal reasoning step
by step:
(1) $P(A\rightarrow C | do(A=1))-P(A\rightarrow C | do(A=0))>0$,
(2) $P(A\rightarrow C | do(B=1))-P(A\rightarrow C | do(B=0))>0$,
(3) $P(A\rightarrow C | do(D=1))-P(A\rightarrow C | do(D=0))=0$,
(4) The causal chains of risk propagation 
is $A \rightarrow B \rightarrow C$. The flood at {\color{blue}\textit{Company A}} disrupted the silk supply to {\color{orange} \textit{Company B}}, which in turn affected {\color{orange}\textit{Company B}}'s ability to supply fabrics to {\color{green}\textit{Company C}}. This sequence of disruptions led to {\color{green}\textit{Company C}}'s delivery delays.  } \\

\bottomrule
\end{tabular}
\begin{tablenotes}
\item[*] 
\small Notably, the nodes in financial KGs precisely correspond to the entities in the text. Here, we visually demonstrate their matching relations through colors.
In the subsequent fusion module, we continue to maintain the correct alignments between such two features.
\end{tablenotes}
\end{threeparttable}
\label{tab:instructions}
\end{table}

\begin{mydef}[\textbf{Hierarchies of Causation}]
Proposed by Pearl and Mackenzie~\cite{pearl2018book}, it is a three-tiered framework for causal reasoning, including association, intervention, and counterfactual.
This framework is pivotal for a deep understanding and accurate inference of causal relationships.

$\bullet$~\textbf{Association:} This focuses on statistical dependencies among random variables, employing probabilistic reasoning regarding joint and conditional distributions, that is $P(X=x, Y=y)$ and $P(Y=y|X=x)$.

$\bullet$~\textbf{Intervention:}
    This allows us to artificially adjust the value of variables (causes) within a causal system, thereby observing the impact of this adjustment on other variables (effects). Because it can control for potential confounding variables, intervention enables us to more directly test and validate causal hypotheses.
    Such interventions can be formalized using the do-operator~\cite{goldszmidt1992rank}, expressed as the distribution of $Y$ being $P(Y=y|do(X=x))$ when setting $X=x$.
    
$\bullet$~\textbf{Counterfactual:} 
    This is a hypothetical way of thinking that explores the impact on outcomes by imagining events as not having occurred or as different from reality, aiming to answer the question, "What would the outcome be if things were different?"
    Counterfactual probabilities
 can be written as $P(Y_x=y)$, representing the probability that "$Y$ would be $y$, had $X$ been $x$".
\end{mydef}

Based on the above hierarchies of causation, we construct two categories of data (as depicted in Table~\ref{tab:instructions}) to block backdoor pathways and fine-tune our model to understand the cause-and-effect relationships behind risk contagion.

\begin{algorithm}[htbp]
    \small
	\caption{Data Intervention Algorithm}
	\label{algo:intervention}
	\LinesNumbered
	\KwIn{Financial knowledge graph (KG), start node \( v_{s} \), target node \( v_{t} \).}
	\KwOut{Set of factual and intervention graphs \( \mathcal{G}_{\text{set}} \).}
	\BlankLine
	
	\( \mathcal{G}_{\text{set}} \leftarrow \emptyset \)\;
    
    \tcp{Perform a depth-first search (DFS) to obtain the factual graph \( \mathcal{G} \) from the financial KG}

    \( \mathcal{G} \leftarrow \text{DFS}(\text{KG}, v_{s}, v_{t}) \)\;
    \( \mathcal{G}_{\text{set}}.\text{add}(\mathcal{G}) \)\;
    
    \tcp{Randomly intervene on each node except \( v_{t} \)}

    \For{each node \( v \) in \( \mathcal{G} \)}{
        \If{ \( v \neq v_{t} \)}{
            \( \widetilde{\mathcal{G}} \leftarrow \text{copy}(\mathcal{G}) \)\;
            Remove \( v \) and its edges in \( \widetilde{\mathcal{G}} \)\;
            \( \mathcal{G}_{\text{set}}.\text{add}(\widetilde{\mathcal{G}}) \)\;
        }
    }
	\BlankLine
	\KwRet{ $\mathcal{G}_{\mathrm{set}}$ }\;
\end{algorithm}

$\bullet$~\textbf{Factual Group:} 
Based on the principle of \textit{Association}, we employ the depth-first search (DFS) algorithm to extract a benchmark factual graph $\mG$ from a financial KG. $\mG$ records the association with regard to risk contagion, including the causal parts $X$ and the non-causal parts $Z$. Without any intervention, the probability of risk contagion is $100\%$.

$\bullet$~\textbf{Intervention Group:}
Following the \textit{Intervention} and \textit{Counterfactual} rule, we introduce random do-operators within $\mG$, following Algorithm~\ref{algo:intervention}. This approach entails directly manipulating specific variables within $\mG$ to validate the effects of these interventions on risk propagation.
For example, the probability of risk contagion is $0\%$ after removing \textit{Company B}.
By comparing with the factual group, we observe a significant change in the probability of risk contagion before and after the intervention on \textit{Company B}. Therefore, we can confirm that \textit{Company B} is a contributing factor to the risk contagion.
In this way, we can clearly distinguish between the causal and non-causal components.

In the explanation, we employ the chain-of-thought approach
to guide LLMs in making causal inferences. As a result, we obtain a causal instruction set $\mI=\{q_i, e_i\}_{i=1}^N$, where $q$ and $e$ represent the query and explanation, respectively.

\subsection{Fusion Module}
\begin{figure}[htbp]
 \includegraphics[width=0.5\textwidth]{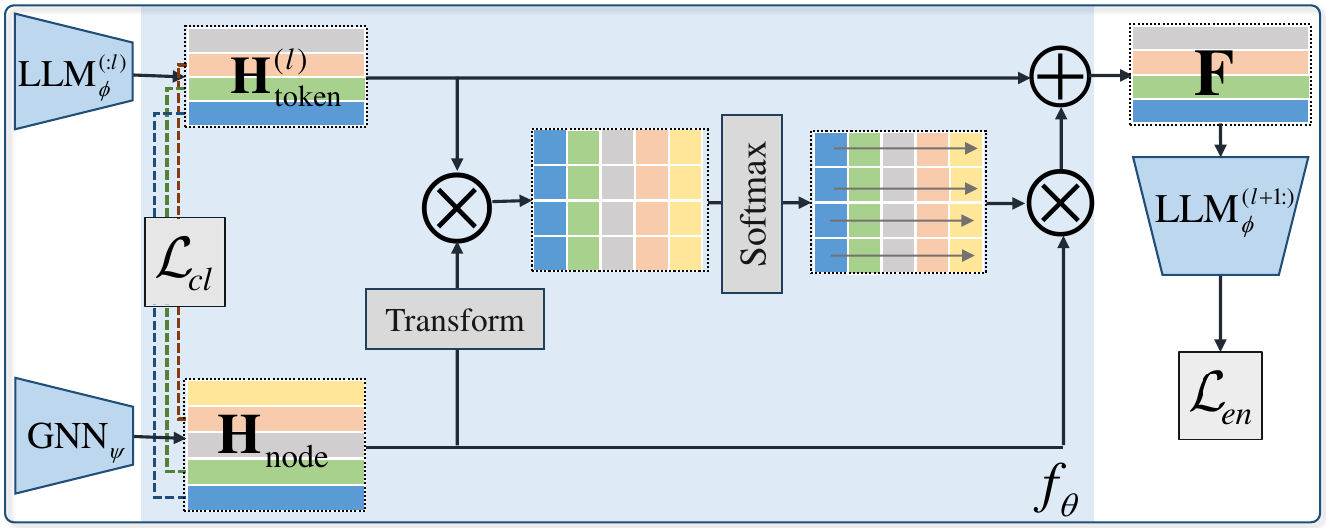}
    \caption{Fusion Module $f_\theta$}
    \label{fig:fusion}
\end{figure}

To integrate LLMs reasoning with KGs at the architectural level, this study introduces a fusion module $f_\theta$, as depicted in Fig.~\ref{fig:fusion}. This module first ensures correct alignments between tokens and nodes via multi-scale contrastive learning. Subsequently, it achieves a deep integration of textual and graph-structured information through cross multi-head attention mechanisms.

\subsubsection{Multi-scale Contrastive Learning for Correct Alignments}

Correct alignments between tokens and nodes
guarantee that $\mathrm{LLM}_\phi$ carries out accurate causal reasoning with KGs. Here, we use a multi-scale contrastive loss function to guide models in learning alignments at varying granularities.

\begin{align}
    \bH_\mathrm{token}^{(l)}&=\mathrm{LLM}^{(:l)}_\phi([q;e])\in\mathbb{R}^{L\times d},\notag\\
    \bH_\mathrm{node} &= \mathrm{GNN}_\psi(\mG_\mathrm{set})\in \mathbb{R}^{M \times d},\notag
\end{align}
where $[q;e]$ denotes the concatenation of queries and explanations with a token delimiter, like $\mathrm{<SEP>}$.
$\mathrm{LLM}^{(:l)}_\phi$ represents
the output of the $l$-th layer of $\mathrm{LLM}_\phi$.
$L$ represents the number of tokens.
$M$ represents the number of nodes in $\mG_\mathrm{set}$.

\begin{align*}
 &\mathcal{L}_\mathrm{token\_node} = \\
 &-\log \frac{\exp\left (\mathds{1}_{[\mathrm{paired}]} \mathrm{sim}(\bH_\mathrm{token}^{(l)}[i], \bH_\mathrm{node}[i])/\tau\right)}{\sum_{i=1}^{\min(L, M)} \mathds{1}_{[\mathrm{ unpaired}]}\exp\left(\text{sim}(\bH_\mathrm{token}^{(l)}[i], \bH_\mathrm{node}[i])/\tau\right)},
\end{align*}

\begin{align*}
     &\mathcal{L}_\mathrm{token\_subgraph}=\\
& -\log \frac{\exp\left (\mathds{1}_{[\mathrm{paired}]} \mathrm{sim}(\bH_\mathrm{token}^{(l)}[i], G_i)/\tau\right)}{\sum_{i=1}^{\min(L,M)} \mathds{1}_{[\mathrm{ unpaired}]}\exp\left(\text{sim}(\bH_\mathrm{token}^{(l)}[i], G_i)/\tau\right)}.
\end{align*}

In the above equation, $\mathcal{L}_\mathrm{token\_node}$ denotes the contrastive loss between tokens and nodes. $\mathcal{L}_\mathrm{token\_subgraph}$ captures the contrastive loss between tokens and $k$-hop subgraphs surrounding the nodes.
$\mathds{1}_{[\mathrm{paired}]}$ serves as a predefined indicator, assigned a value of $1$ if the $i$-th token and the $i$-th node are matched, and $0$ otherwise.
For example, if the token position and node position corresponding to \textit{Company A} are the same, then $\mathds{1}_{[\mathrm{paired}]}$ is assigned a value of $1$.
$\mathds{1}_{[\mathrm{unpaired}]}$ represents the complementary indicator.
$\bH_\mathrm{token}^{(l)}[i]$ and $\bH_\mathrm{node}[i]$ correspond to the embeddings of the $i$-th token and the $i$-th node, respectively.
$\tau$ is the temperature parameter that modulates the scale of similarity scores between tokens and nodes, influencing the smoothness of the resulting probability distribution. A lower $\tau$ value emphasizes minor score discrepancies, leading to a more concentrated distribution, whereas a higher value yields a more uniform distribution.
$G_i$ represents the embedding of the $k$-hop subgraphs associated with the $i$-th node.
By considering the $w$-span context around the $i$-th token, we construct $\mathcal{L}_\mathrm{context\_node}$ and $\mathcal{L}_\mathrm{context\_subgraph}$.
The following equation is the multi-scale contrastive loss function that effectively captures relationships at varying granularities, enhancing the model's ability to discern and leverage hierarchical information between tokens and nodes.

\begin{align}
\begin{aligned}
    \mathcal{L}_{cl} &=  \mathcal{L}_\mathrm{token\_node} + \mathcal{L}_\mathrm{token\_subgraph}\\ 
    &+\mathcal{L}_\mathrm{context\_node}+\mathcal{L}_\mathrm{context\_subgraph}.
\end{aligned}
\label{eq:cl}
\end{align}

\subsubsection{Soft Prompt with Cross Multi-head Attention Mechanisms}
Here, we employ a multi-head attention mechanism with $J$ heads to learn the information fusion between tokens and nodes.
\begin{align}
    \bQ_j &= \text{Q-Linear}_j(\bH_\mathrm{token}^{(l)}) \in \mathbb{R}^{L\times \frac{d}{J}},\\
    \bK_j &= \text{K-Linear}_j(\bH_\mathrm{node}) \in \mathbb{R}^{M\times \frac{d}{J}},\\
    \bV_j &= \text{V-Linear}_j(\bH_\mathrm{node}) \in \mathbb{R}^{M\times \frac{d}{J}},\\
    \bA_j &= \mathrm{Softmax}(\frac{
    \bQ_j\bK_j^\top}{\sqrt{d/J}})\in \mathbb{R}^{L\times M},\\
    \bP_j &= \bA_j \bV_j\in \mathds{R}^{L\times \frac{d}{J}},\\
    \bP&=[\bP_1\|\bP_2\|\cdots\|\bP_J]\in \mathbb{R}^{L\times d},\\
    \bF & =  [\bP \oplus \bH_\mathrm{token}^{(l)}]\in \mathbb{R}^{2L\times d},\\
    \bT &= \mathrm{LLM}^{(l+1:)}_\phi(\bF) \in \mathbb{R}^{L\times C},\label{eq:alpha}\\
    \mathcal{L}_{en} &= -\frac{1}{L}\sum_{\iota=1}^{L}\sum_{c=1}^{C} \bI_{\iota,c} \log(\bT_{\iota,c}[L+1:]).
\end{align}

In the equations presented, the linear transformation $\text{Q-Linear}_j(\cdot)$ for the $j$-th attention head is initially applied to the token representations from the $l$-th layer of $\mathrm{LLM}_\phi$, producing query matrices $\bQ_j$. Simultaneously, node representations are transformed linearly to create key matrices $\bK_j$ and value matrices $\bV_j$.
Following this, the attention weight matrix $\bA_j$ is calculated from the product of $\bQ_j$ and $\bK_j$, with the application of the $\mathrm{Softmax}(\cdot)$ function. This weight matrix, together with $\bV_j$, generates the output $\bP_j$ for each attention head.
The outputs $\{\bP_j\}_{j=1}^J$ are then concatenated to form the soft prompt $\bP$. 
The concatenation of $\bP$ with $\bH_\mathrm{token}^{(l)}$
along the sequence dimension results in $\bF$.
This combined output is subsequently processed through $\mathrm{LLM}_\phi$, leading to the token
probabilities $\bT$. 
Finally, we optimize our model using the cross-entropy loss function, where $\bI$ represents the ground-truth token labels.
$\bT[L+1:]$ represents the predicted probabilities of
the last $L$ tokens. $C$ represents the vocabulary size.

\subsection{Risk Pathway Inference Module }
To infer the paths of risk propagation, we further conduct the following risk pathway inference module for calculating
risk scores for each node in the graph. 

\begin{align*}
    &\widehat{P}(v_s\rightarrow 
    v_t|do(\nu=1~or~0))  = \\ &\mathrm{Sigmoid}\left(\mathrm{Readout}( \text{ S-Linear}([\bA_1, \bA_2,\cdots, \bA_J]^\top)) \right),
\end{align*}

\begin{align*}
    \mathcal{L}_{X} = \mathbb{E}_{\nu \in X}&[\widehat{P}(v_s\rightarrow 
    v_t|do(\nu=1))\\&-\widehat{P}(v_s\rightarrow v_t|do(\nu=0))],\\
    \mathcal{L}_{Z} = \mathbb{E}_{\nu\in Z}&[\widehat{P}(v_s\rightarrow v_t|do(\nu=1))\\&-\widehat{P}(v_s\rightarrow v_t|do(\nu=0))],\\
    \mathcal{L}_{path} = \mathcal{L}_{Z}&-\mathcal{L}_{X}.
\end{align*}

In the above equations, \( \widehat{P}(v_s\rightarrow 
    v_t|do(\nu=1~or~0))  \) denotes the estimated probabilities of risk contagion from the initiating node \( v_s \) to the target node \( v_t \).
The $\text{ S-Linear}(\cdot)$ function maps a high-dimensional
matrix into a $1$-D space. 
The $\mathrm{Readout}(\cdot)$ function is responsible for aggregating node-level features into graph-level features by computing the average of the node features.
The variables \( X \) and \( Z \) signify causal and non-causal nodes, respectively.
The notation \( do(\nu = 0) \) signifies an intervention applied to node \( \nu \), effectively setting its state to zero.
Our objective is to achieve a state where \( \widehat{P} \) equals 0 upon intervention at causal nodes, and where \( \widehat{P} \) remains 1 when intervening at non-causal nodes.

\subsection{Joint Tuning}
\begin{figure}[htbp]
\centering 
    \includegraphics[width=1\linewidth]{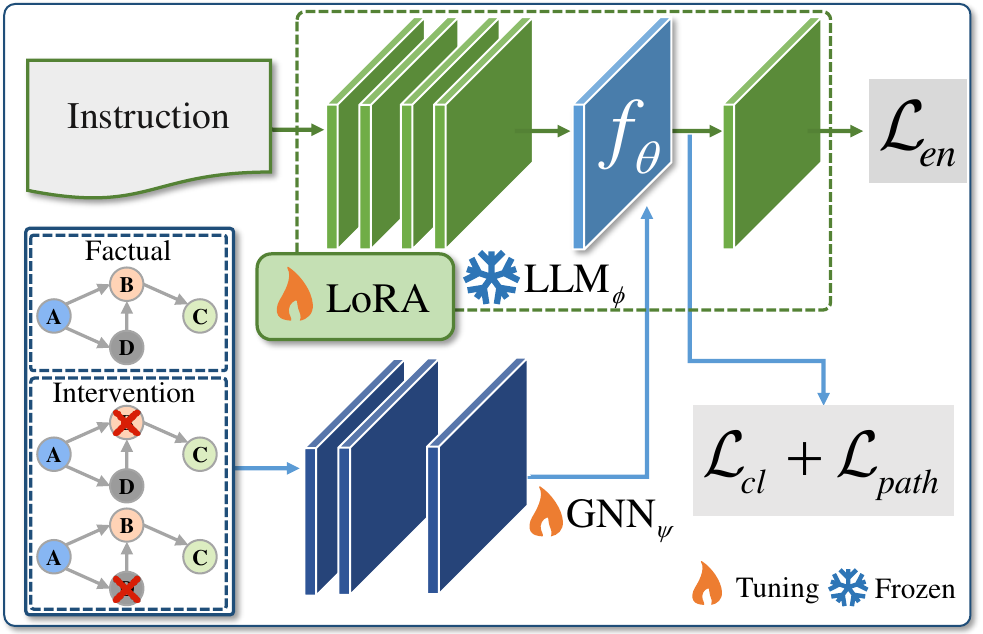}
    \caption{Joint Fine-tuning of Our Proposed RC\textsuperscript{2}R}
    \label{fig:tuning}
\end{figure}

To enable RC\textsuperscript{2}R to effectively learn the causal mechanisms in financial risk propagation, we adopt the fine-tuning mechanism illustrated in Fig.~\ref{fig:tuning}.
$N$ pairs of instructions $\mI$ and financial KGs $\mG$ are input into $\mathrm{LLM}_\phi$ and $\mathrm{GNN}_\psi$, respectively. To train the overall model in a unified manner, we formulate the following joint loss function,
\vspace{-2mm}
\begin{align}
    \mathcal{L}_{joint} = \mathcal{L}_{cl} + \mathcal{L}_{en}+\mathcal{L}_{path}.
\end{align}
Minimizing this loss function is designed to ensure precise alignments between textual and graph-based information, while also enabling LLMs to accurately generate explanations.

\section{Experimental Evaluation}
In this section, we conduct a series of experiments on two open-sourced datasets to validate the performance of our proposed RC\textsuperscript{2}R. We compare the effectiveness of predicted risk contagion and the quality of generated explanations. Additionally, we report and visualize the reasoning results of risk contagion pathways.
Ultimately, we conduct ablation studies to verify the functionality and significance of each component within our proposed model.

\subsection{Datasets}
\begin{table}[htbp]
\small
\centering
\caption{Percentage of Financial Topics in Two Datasets}
\begin{tabular}{cccc}
\toprule
Datasets     & Topics & Percentages \\
\midrule
\multirow{5}{*}{FinDKG} &  Stock & 35\% \\
             & Bond         & 25\%       \\
             & Money         & 20\%       \\
             & Real estate  & 10\%       \\
             & Commodity     & 10\%       \\
\midrule
\multirow{4}{*}{SupplyChain-KG} & Raw materials     &20\% \\
& Manufacturing & 30\%\\
& Wholesale & 20\%\\
& Retail & 30\%\\
\bottomrule
\label{tab:data description}
\end{tabular}
\end{table}

We open-source two comprehensive datasets, FinDKG and SupplyChain-KG, for the analysis of financial risk contagion. 
Each dataset encompasses causal texts and financial KGs.

\hspace{1em}$\bullet$~\textbf{FinDKG}~\cite{li2023findkg}: 
This dataset covers $15$ entity types and $15$ relation types, encompassing a total of $13,645$ nodes and $242,149$ edges. Based on this open-source dataset, we have constructed $5,000$ causal instructions, covering diverse markets such as stock, bond, money, real estate, and commodity. The data ratios for each market are shown in Table~\ref{tab:data description}. 
Following the guidelines in Table~\ref{tab:instructions}, we have designed tasks for each market, comprising $15\%$ factual groups and $85\%$ intervention groups.

$\bullet$~\textbf{SupplyChain-KG}: 
This dataset encompasses $10$ entity types and $40$ relation types, totaling $2,400$ nodes and $136,005$ edges. We have developed $5,000$ causal instructions addressing various risks within the supply chain, including financial shortages, natural and man-made disasters, and market volatility. Data distribution across different markets is detailed in Table~\ref{tab:data description}. Following the guidelines in Table~\ref{tab:instructions}, we have crafted tasks for each supply chain sector, consisting of $17\%$ factual groups and $83\%$ intervention groups.

\subsection{Experimental Settings}
\subsubsection{Baseline Models}
Here, we provide the following four categories of baseline models, including LLMs, LLMs+KGs, GNNs, and $\text{GNN}_{cause}$.

$\bullet$~\textbf{LLMs}: \textbf{Gemma-7B} \cite{gemma2024}, which possesses a total parameter size of approximately $7$ billion. The model is composed of $28$ Transformer decoders, each with a hidden size of $3,072$ and $16$ attention heads. Additionally, the feedforward hidden dimension size for this model is $49,152$. The vocabulary size is set at $256,128$.
\textbf{InternLM2-7B}~\cite{team2023internlm}, which possesses a total parameter size of approximately $7$ billion, has demonstrated strong capabilities in text generation and comprehension. The model is composed of $32$ Transformer decoder layers, each with a hidden size of $4,096$ and $32$ attention heads. The vocabulary size is set at $92,544$.
 \textbf{Llama2-7B}~\cite{touvron2023llama}, is a pretrained generative text model with scales of $7$ billion parameters. The model is composed of $32$ Transformer blocks, each with a hidden size of $4,096$ and $32$ attention heads. The vocabulary size is set at $32,000$.
The carefully designed prompts for LLMs are reflected in Table~\ref{tab:prompt}.

$\bullet$~\textbf{LLMs+KGs}: $\textbf{Gemma+}t_\mG$ is a combined framework of Gemma and KG information. 
In order to contextualize the large language model, we transform the KG information into triplet text $t_\mG$ that serves as a context for model input.
The elaborated prompts for LLMs+KGs are reflected in Table~\ref{tab:prompt}.

$\bullet$~\textbf{GNNs}:
The Graph Convolutional Network (\textbf{GCN})~\cite{kipf2016semi} applies convolutional operations on the nodes of a graph, enabling each node to aggregate information from its neighboring nodes, thereby learning the representation of each node.
Graph Attention Network (\textbf{GAT})~\cite{velivckovic2018graph} leverages attention mechanisms to weigh the importance of neighboring nodes in graph-structured data, enabling feature aggregation with varying emphasis on different neighbors.

$\bullet$~$\textbf{GNN}_{cause}$:
The Discovering Invariant Rationale (\textbf{DIR}) method~\cite{wu2021discovering} aims to uncover causal rationales that remain consistent across diverse distributions, thereby facilitating the development of inherently interpretable GNNs.
%
\textbf{V-REx}~\cite{krueger2021out} represents a streamlined adaptation of risk extrapolation. It effectively minimizes risk disparities across training environments, and diminishes model vulnerability to extensive distributional shifts.

\begin{table}[htbp]
    \centering
    \caption{Prompts Used for LLMs \& LLMs+KGs}
    \label{tab:prompt}
    \begin{tabular}{>{\centering\arraybackslash}m{1.5cm}  >{\arraybackslash}m{6cm}}
    \toprule
    
         Role & You are a powerful causal reasoning assistant. \\
        \hline
         Triples $t_\mG$ & [Company A, partners with, Company B], ..., [Company E, supplies to, Company C] \\
        \hline
         Query & Company A produces silk, ..., Company C produces qipao. In Jan, Company A was at risk. In May, Company C was at risk. Please infer the risk contagion pathways from Company A to Company C. \\
         
    \bottomrule

    \end{tabular}
     \begin{threeparttable}
    \begin{tablenotes}
   
    \item[*] 
    \small Notably, the LLMs+KGs baseline models utilize prompts that include role, triples, and query, whereas the LLMs baseline models employ prompts comprising only role and query.
\end{tablenotes}
\end{threeparttable}
\end{table}

\begin{table*}[h]
\centering
\caption{ACC \& AUC of Predicted Risk Contagion}
\begin{threeparttable}
\begin{tabular}{cc|cc|cc} 
\toprule
& Dataset      & \multicolumn{2}{c|}{FinDKG} & \multicolumn{2}{c}{SupplyChain-KG}  \\ 
\midrule
Category & Metric       & ACC   & AUC                 & ACC & AUC                         \\ 
\midrule
\multirow{4}{*}{LLMs+KGs} & \textbf{RC\textsuperscript{2}R (Our Model)}   & $\mathbf{0.783}\pm{0.037}$ & $\mathbf{0.757}\pm{0.045}$ & $\mathbf{0.667}\pm{0.021}$ & $\mathbf{0.630}\pm{0.030}$ \\
& RC\textsuperscript{2}R$^\dagger$   & $0.710\pm{0.028}$ & $0.671\pm{0.033}$ & $0.622\pm{0.024}$ & $0.610\pm{0.019}$ \\
& RC\textsuperscript{2}R$^\ddagger$   & $0.717\pm{0.041}$ & $0.697\pm{0.052}$ & $0.632\pm{0.029}$ & $0.619\pm{0.031}$ \\
& Gemma+$t_\mG$   & $0.747\pm{0.022}$ & $0.671\pm{0.044}$ & $0.625\pm{0.036}$ & $0.611\pm{0.027}$ \\
\midrule
\multirow{3}{*}{LLMs} & Gemma-7B~\cite{gemma2024} & $0.667\pm{0.031}$ & $0.549\pm{0.039}$ & $0.603\pm{0.028}$ & $0.584\pm{0.024}$ \\
& InternLM2-7B~\cite{team2023internlm} & $0.663\pm{0.025}$ & $0.608\pm{0.049}$ & $0.592\pm{0.037}$ & $0.621\pm{0.053}$ \\
& Llama2-7B~\cite{touvron2023llama}  & $0.654\pm{0.023}$ & $0.620\pm{0.035}$ & $0.587\pm{0.031}$ & $0.546\pm{0.029}$ \\
\midrule
\multirow{2}{*}{GNNs} & GCN~\cite{kipf2016semi} & $0.633\pm{0.027}$ & $0.579\pm{0.023}$ & $0.565\pm{0.036}$ & $0.524\pm{0.022}$ \\
& GAT~\cite{velivckovic2018graph} & $0.647\pm{0.041}$ & $0.596\pm{0.033}$ & $0.583\pm{0.029}$ & $0.523\pm{0.018}$ \\
\midrule
\multirow{2}{*}{$\text{GNN}_{cause}$} & DIR~\cite{wu2021discovering} & $0.642\pm{0.038}$ & $0.571\pm{0.028}$ & $0.562\pm{0.025}$ & $0.521\pm{0.023}$ \\
& V-REx~\cite{krueger2021out} & $0.672\pm{0.044}$ & $0.601\pm{0.039}$ & $0.573\pm{0.032}$ & $0.543\pm{0.031}$ \\
\bottomrule
\end{tabular}
\end{threeparttable}
\label{tab:answers}
\end{table*}

\begin{table*}[h]
\centering
\caption{ACC \& AUC of Predicted Risk Contagion on Out-of-distribution Data}
\begin{threeparttable}
\begin{tabular}{cc|>{\centering\arraybackslash}p{2cm}>{\centering\arraybackslash}p{2cm}|>{\centering\arraybackslash}p{2cm}>{\centering\arraybackslash}p{2cm}} 
\toprule
& Training$\rightarrow$Testing& \multicolumn{2}{c|}{SupplyChain-KG$\rightarrow$FinDKG} & \multicolumn{2}{c}{FinDKG$\rightarrow$SupplyChain-KG}  \\ 
\midrule
Category & Metric       & ACC   & AUC                 & ACC & AUC                         \\ 
\midrule
\multirow{4}{*}{LLMs+KGs} & \textbf{RC\textsuperscript{2}R (Our Model)}   & $\mathbf{0.762}\pm{0.021}$ & $\mathbf{0.721}\pm{0.055}$ & $\mathbf{0.653}\pm{0.019}$ & $\mathbf{0.610}\pm{0.034}$ \\
& RC\textsuperscript{2}R$^\dagger$   & $0.695\pm{0.014}$ & $0.654\pm{0.049}$ & $0.604\pm{0.026}$ & $0.591\pm{0.013}$ \\
& RC\textsuperscript{2}R$^\ddagger$   & $0.711\pm{0.017}$ & $0.677\pm{0.033}$ & $0.608\pm{0.027}$ & $0.581\pm{0.022}$ \\
& Gemma+$t_\mG$   & $0.725\pm{0.018}$ & $0.683\pm{0.042}$ & $0.612\pm{0.021}$ & $0.596\pm{0.013}$ \\
\midrule
\multirow{3}{*}{LLMs} & Gemma-7B~\cite{gemma2024} & $0.611\pm{0.028}$ & $0.560\pm{0.045}$ & $0.600\pm{0.022}$ & $0.573\pm{0.019}$ \\
& InternLM2-7B~\cite{team2023internlm} & $0.612\pm{0.035}$ & $0.605\pm{0.021}$ & $0.581\pm{0.044}$ & $0.548\pm{0.032}$ \\
& Llama2-7B~\cite{touvron2023llama}  & $0.608\pm{0.039}$ & $0.580\pm{0.025}$ & $0.582\pm{0.033}$ & $0.541\pm{0.018}$ \\
\midrule
\multirow{2}{*}{GNNs} & GCN~\cite{kipf2016semi} & $0.553\pm{0.048}$ & $0.531\pm{0.015}$ & $0.584\pm{0.042}$ & $0.554\pm{0.011}$ \\
& GAT~\cite{velivckovic2018graph} & $0.610\pm{0.031}$ & $0.547\pm{0.039}$ & $0.578\pm{0.047}$ & $0.553\pm{0.027}$ \\
\midrule
\multirow{2}{*}{$\text{GNN}_{cause}$} & DIR~\cite{wu2021discovering} & $0.627\pm{0.036}$ & $0.568\pm{0.044}$ & $0.533\pm{0.015}$ & $0.518\pm{0.034}$ \\
& V-REx~\cite{krueger2021out} & $0.619\pm{0.029}$ & $0.597\pm{0.022}$ & $0.558\pm{0.031}$ & $0.533\pm{0.026}$ \\
\bottomrule
\end{tabular}
\end{threeparttable}
\label{tab:ood-answers}
\end{table*}

\subsubsection{Parameter Configuration of Our Model}

According to our preliminary experiments, we choose the following parameters to ensure optimal performance. 
Our selection of LLMs includes the Gemma-7B~\cite{gemma2024}, specifically fine-tuned via the LoRA method~\cite{hu2021lora}, to achieve enhanced model adaptability and performance.
For GNNs, we prioritize models incorporating residual connections, notably the GCN~\cite{kipf2016semi}. 
The neural network architecture consists of $4$ layers, a configuration that balances complexity with performance efficacy. 
We standardize the dimension of hidden layers at $1024$, a decision guided by initial tests.

In multi-scale contrastive learning, we set $k=2$ for subgraph extraction, $w=3$ for context selection, and set $\tau$ to 1.
Our configuration of the cross-attention mechanisms includes $8$ heads.
A batch size of $1$ is selected, an unconventional choice that our empirical evidence suggests maximizes training efficiency and model convergence.
The training spans $5$ epochs, with a learning rate set at $0.01$.
Furthermore, to assess the effectiveness of cross attention mechanisms and multi-scale contrastive learning, we develop two variants: $\textbf{RC\textsuperscript{2}R}^\dagger$, which replaces cross attention mechanisms with concatenation, and $\textbf{RC\textsuperscript{2}R}^\ddagger$, which omits multi-scale contrastive learning.

\subsubsection{Evaluation Metrics}
To evaluate the effectiveness of predicted risk contagion, we measure them using the accuracy (\textbf{ACC}) and area under the curve (\textbf{AUC}) metrics.
%

To assess the quality of generated explanations, we consider the following five key criteria. (1) \textbf{Fluency} measures the naturalness and readability, determining if the text flows smoothly. (2) \textbf{Relevance} examines how well the text aligns with the given task or query. (3) \textbf{Consistency} checks for any contradictions within the information presented. (4) \textbf{Diversity} evaluates the range of different yet relevant outputs the model can produce. (5) \textbf{Informativeness} gauges the richness of useful information within the text.
Additionally, we will report the average scores ($[0, 10]$) for each criterion based on both human evaluation and GPT-4 assessment, with higher scores indicating better performance.

To assess the predictive accuracy of risk propagation pathways, we employ the intersection over union (\textbf{IoU}) metric, defined as $\text{IoU} = \frac{\text{path intersection}}{\text{path union}}\in[0, 1]$. Here, the intersection denotes the count of nodes that overlap between the predicted and ground-truth paths, whereas the union represents the aggregate count of unique nodes obtained by combining those from both the predicted paths and the ground-truth paths. A higher $\text{IoU}$ indicates better performance.

\subsubsection{Experimental Platform}
All experiments are conducted on a Linux server with a GPU (NVIDIA A800, Memory 80G) and CPU (Hygon C86 7375 32-core Processor). 
We implement our proposed model with deep learning library PyTorch, transformers, and Deep Graph Library (dgl). The versions of Python, PyTorch, transformers, and dgl are 3.8.10, 2.2.0+cu118, 4.38.1, 1.1.2+cu118, respectively.

\subsection{Experimental Results \& Analysis}
In this section, we evaluate the performance of our proposed RC\textsuperscript{2}R from three perspectives: predicted risk contagion, generated explanations, and inferred propagation pathways.

\subsubsection{Performance of Predicted Risk Contagion}

From Tables~\ref{tab:answers} and \ref{tab:ood-answers}, we observe that RC\textsuperscript{2}R outperforms Gemma+$t_\mG$, achieving an average increase of $3.9\%$ in ACC and $5.25\%$ in AUC. This improvement is attributed to the ability of our model to effectively leverage the structural information in KGs via GNNs, whereas Gemma+$t_{\mG}$ destroys the spatial structure of graphs by converting them into triplet texts.
When compared to LLMs, RC\textsuperscript{2}R demonstrates a significant average improvement of approximately $9.733\%$ in ACC and $10.55\%$ in AUC. This is largely due to the proficiency of our model in extracting crucial background information from KGs for causal reasoning.
Against GNNs, our model shows a remarkable average improvement of $11.8\%$ in ACC and $13.8\%$ in AUC. 
This boost is a result of the successful integration of the semantic understanding and logical reasoning capabilities of LLMs.
Furthermore, compared to $\text{GNN}_{cause}$, our model achieves an average improvement of $11.275\%$ in ACC and $13.45\%$ in AUC. 
This progress is due to the extensive causal instructions for model tuning, effectively guiding our model in conducting causal reasoning on risk propagation.

\subsubsection{Quality of Generated Explanations}

To assess the quality of explanations generated by RC\textsuperscript{2}R, we combine human evaluation (by 8 evaluators) and GPT-4 review to compare our model with four baseline models. Each model is evaluated using 8 random samples. All evaluation results are ultimately aggregated into average values and presented in Tables \ref{tab:explanation-1} and \ref{tab:explanation-2}.
From them, we find that our model excels in several key indicators of explanation quality, especially in diversity, informativeness, and consistency. 
Specifically, our model shows significant improvements in terms of diversity, with an increase of 1.25. Similarly, it achieves a notable enhancement in informativeness, with an approximate increase of 0.28. 
Such advancements are primarily attributed to the integration of knowledge content from the KGs, which enriches the explanatory depth of our model's output. Concurrently, there is an improvement in consistency, around 0.21, mainly due to contrastive learning facilitating effective alignment between text and nodes, ensuring that the content generated by the model remains consistent with the posed queries.

\subsubsection{Performance of Inferred Propagation Pathways}
\begin{figure}[htbp]
\centering 
    \includegraphics[width=0.5\textwidth]{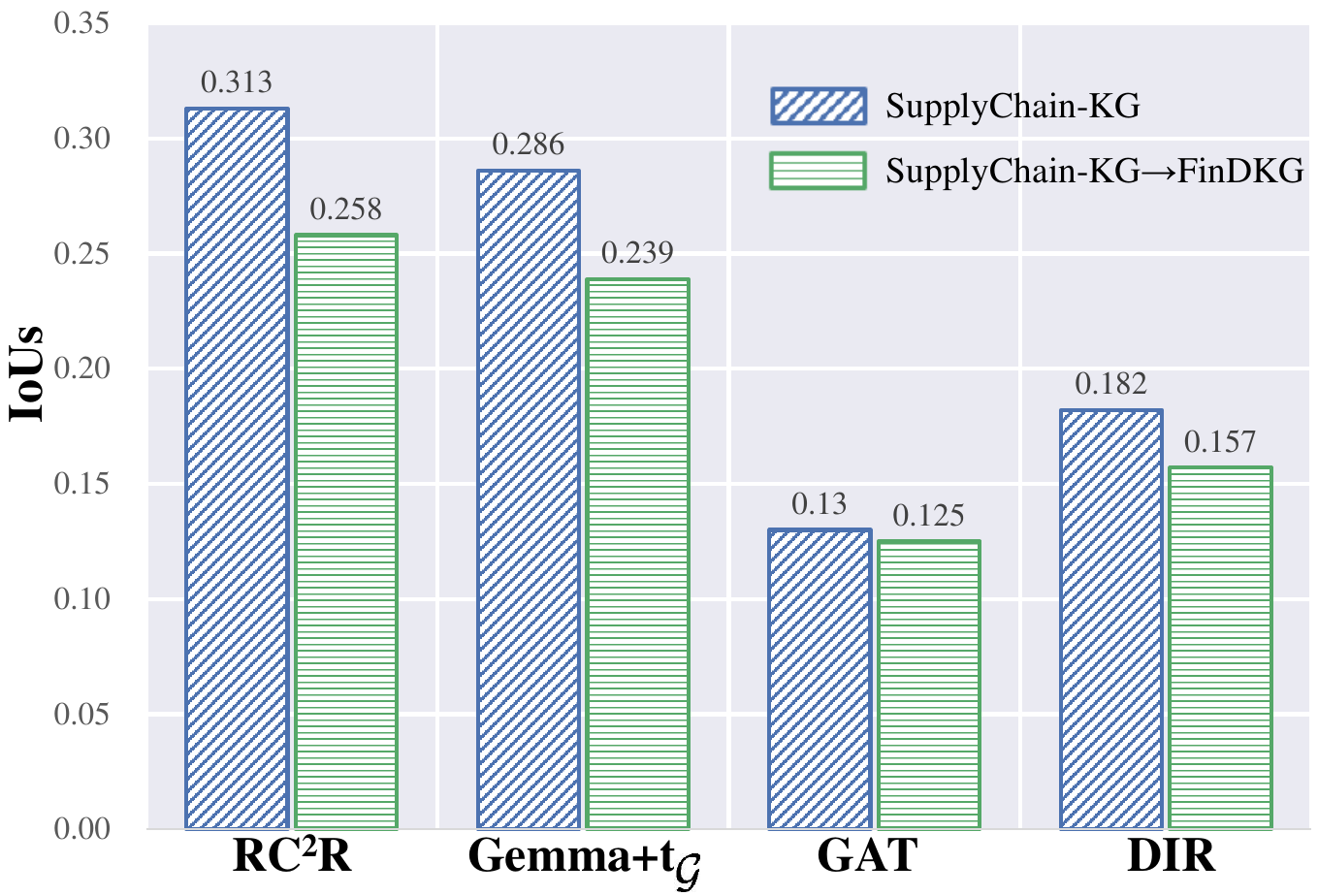}
    \caption{IoUs of Inferred Propagation Pathways}
    \label{fig:IoUs}
\end{figure}

In Fig.~\ref{fig:IoUs}, RC\textsuperscript{2}R outperforms Gemma+$t_\mG$ with a $2.3\%$ increase in the IoU score. This gain is linked to our model's effective integration of graph information, enhancing its ability to accurately identify pathways for inference propagation.
Compared to GAT, our model secures a $15.8\%$ improvement in the IoU score. This significant boost stems from the model's skillful leverage of inferential capabilities inherent in LLMs, moving beyond the exclusive reliance on the strengths of GNNs.
When measured against DIR, our model shows an $11.6\%$ increase in the IoU score.
This enhancement primarily stems from our model's ability to parse textual semantics and comprehensively grasp various factual and intervention groups within causal instructions.

\begin{table*}[htbp]
\centering
\setlength{\tabcolsep}{2pt} %
\caption{Quality of Generated Explanations}
\begin{threeparttable}
\begin{tabular}{c|c|c|c|c|c|c|c|c|c|c} 
\toprule
Metrics& \multicolumn{2}{c|}{Diversity} & \multicolumn{2}{c|}{Informativeness} & \multicolumn{2}{c|}{Consistency} &\multicolumn{2}{c|}{Relevance} & \multicolumn{2}{c}{Fluency} \\ 
\midrule
Experts& Human & GPT-4 & Human & GPT-4   & Human & GPT-4  & Human & GPT-4   & Human & GPT-4 \\ 
\midrule
\textbf{RC\textsuperscript{2}R (Our Model)} & $\mathbf{5.38} \pm 0.60$  & $\mathbf{4.50} \pm 0.20$   &  $\mathbf{5.50} \pm 0.55$     & $6.25 \pm 0.15$     &  $7.25 \pm 0.45$ & $8.50 \pm 0.16$ & $\mathbf{8.00} \pm 0.75$      &  $8.63 \pm 0.15$ &$8.13 \pm 0.50$ & $8.38 \pm 0.05$   \\
Gemma+$t_\mG$  &  $5.13 \pm 0.75$    &  $4.00 \pm 0.20$   &  $5.13 \pm 0.55$  &  $6.00 \pm 0.20$    & $7.00 \pm 0.25$  & $8.88 \pm 0.05$ & $8.00 \pm 0.40$   &  $9.00 \pm 0.13$ & $8.63 \pm 0.05$  & $9.25 \pm 0.10$    \\
Gemma-7B~\cite{gemma2024} &   $3.00 \pm 0.55$    &  $2.63 \pm 0.32$    &  $4.88 \pm 0.24$&  $6.13 \pm 0.05$      &  $6.50 \pm 0.25$       & $8.50 \pm 0.18$  &  $7.88 \pm 0.47$  & $8.50 \pm 0.15$  & $8.00 \pm 0.50$  & $9.00 \pm 0.05$     \\
InternLM2-7B~\cite{team2023internlm} &  $4.63 \pm 0.45$ &  $4.00 \pm 0.06$   &  $5.00 \pm 0.55$ & $7.00 \pm 0.23$    &  $7.38 \pm 0.60$ &  $8.00 \pm 0.12$ & $7.38 \pm 0.48$  &  $8.50 \pm 0.12$ & $8.13 \pm 0.62$ & $9.00 \pm 0.10$     \\
Llama2-7B~\cite{touvron2023llama} &  $3.00 \pm 0.40$  &  $3.13 \pm 0.20$ &  $4.50 \pm 0.52$ & $6.13 \pm 0.11$     &  $6.88 \pm 0.32$ &  $8.13 \pm 0.08$ & $7.38 \pm 0.48$  &  $9.00 \pm 0.05$ & $8.00 \pm 0.50$ & $8.50 \pm 0.15$  \\
\midrule
Count & \multicolumn{2}{c|}{$8$} & \multicolumn{2}{c|}{$7$} &  \multicolumn{2}{c|}{$5$} & \multicolumn{2}{c|}{$5$} & \multicolumn{2}{c}{$2$} \\ 
\bottomrule
\end{tabular}
\begin{tablenotes}
\item[*] Count denotes the frequency at which our method outperforms other methods.
\end{tablenotes}
\end{threeparttable}
\label{tab:explanation-1}
\end{table*}

\begin{table*}[htbp]
\centering
\setlength{\tabcolsep}{2pt} 
\caption{Quality of Generated Explanations on Out-of-distribution Data (SupplyChain-KG$\rightarrow$FinDKG)}
\begin{tabular}{c|c|c|c|c|c|c|c|c|c|c} 
\toprule
Metrics& \multicolumn{2}{c|}{Diversity} & \multicolumn{2}{c|}{Informativeness} & \multicolumn{2}{c|}{Consistency} &\multicolumn{2}{c|}{Relevance} & \multicolumn{2}{c}{Fluency} \\ 
\midrule
Experts & Human & GPT-4& Human & GPT-4  & Human & GPT-4  & Human & GPT-4 & Human & GPT-4 \\ 
\midrule
\textbf{RC\textsuperscript{2}R  (Our Model)} &  $\mathbf{5.38} \pm 0.32$   & $4.13 \pm 0.17$   &  $\mathbf{5.13} \pm 0.25$ &   $\mathbf{6.00} \pm 0.10$  &  $\mathbf{7.25} \pm 0.60$ & $7.50 \pm 0.22$ &  $7.63 \pm 0.60$     &  $7.50 \pm 0.08$ &  $8.38 \pm 0.48$ &  $8.50 \pm 0.18$    \\
Gemma+$t_\mG$ & $4.50 \pm 0.25$ & $3.88 \pm 0.04$   & $4.50 \pm 0.15$ &   $5.50 \pm 0.05$ &  $6.63 \pm 0.32$ & $7.25 \pm 0.12$ & $8.13 \pm 0.46$  &    $9.13 \pm 0.19$  &  $8.50 \pm 0.52$ & $9.50 \pm 0.20$    \\
Gemma-7B~\cite{gemma2024} &  $3.00 \pm 0.20$ & $3.00 \pm 0.02$     &   $3.88 \pm 0.15$   &   $3.13 \pm 0.12$ &  $6.00 \pm 0.22$     &  $7.25 \pm 0.15$  &  $7.50 \pm 0.45$  &  $8.50 \pm 0.20$ & $8.38 \pm 0.40$  & $9.00 \pm 0.09$         \\
InternLM2-7B~\cite{team2023internlm} & $5.00 \pm 0.22$  & $4.38 \pm 0.04$  & $4.88 \pm 0.10$  &   $5.63 \pm 0.14$  & $6.50 \pm 0.40$  & $8.00 \pm 0.13$  & $7.13 \pm 0.32$ & $8.00 \pm 0.05$ &  $8.13 \pm 0.33$ & $8.88 \pm 0.16$    \\
Llama2-7B~\cite{touvron2023llama} & $3.50 \pm 0.13$   &  $3.00 \pm 0.07$ & $4.50 \pm 0.15$ &  $5.63 \pm 0.03$ & $6.50 \pm 0.22$  &  $7.00 \pm 0.10$ &$6.88 \pm 0.33$  & $7.13 \pm 0.03$ &  $8.13 \pm 0.45$    & $8.50 \pm 0.20$     \\
\midrule
Count &  \multicolumn{2}{c|}{$7$} & \multicolumn{2}{c|}{$8$} & \multicolumn{2}{c|}{$7$} & \multicolumn{2}{c|}{$4$} & \multicolumn{2}{c}{$2$} \\ 
\bottomrule
\end{tabular}
\label{tab:explanation-2}
\end{table*}

\subsubsection{Visualization of Propagation Pathways}
\begin{figure*}[htbp]
\centering 
    \includegraphics[width=0.8\textwidth]{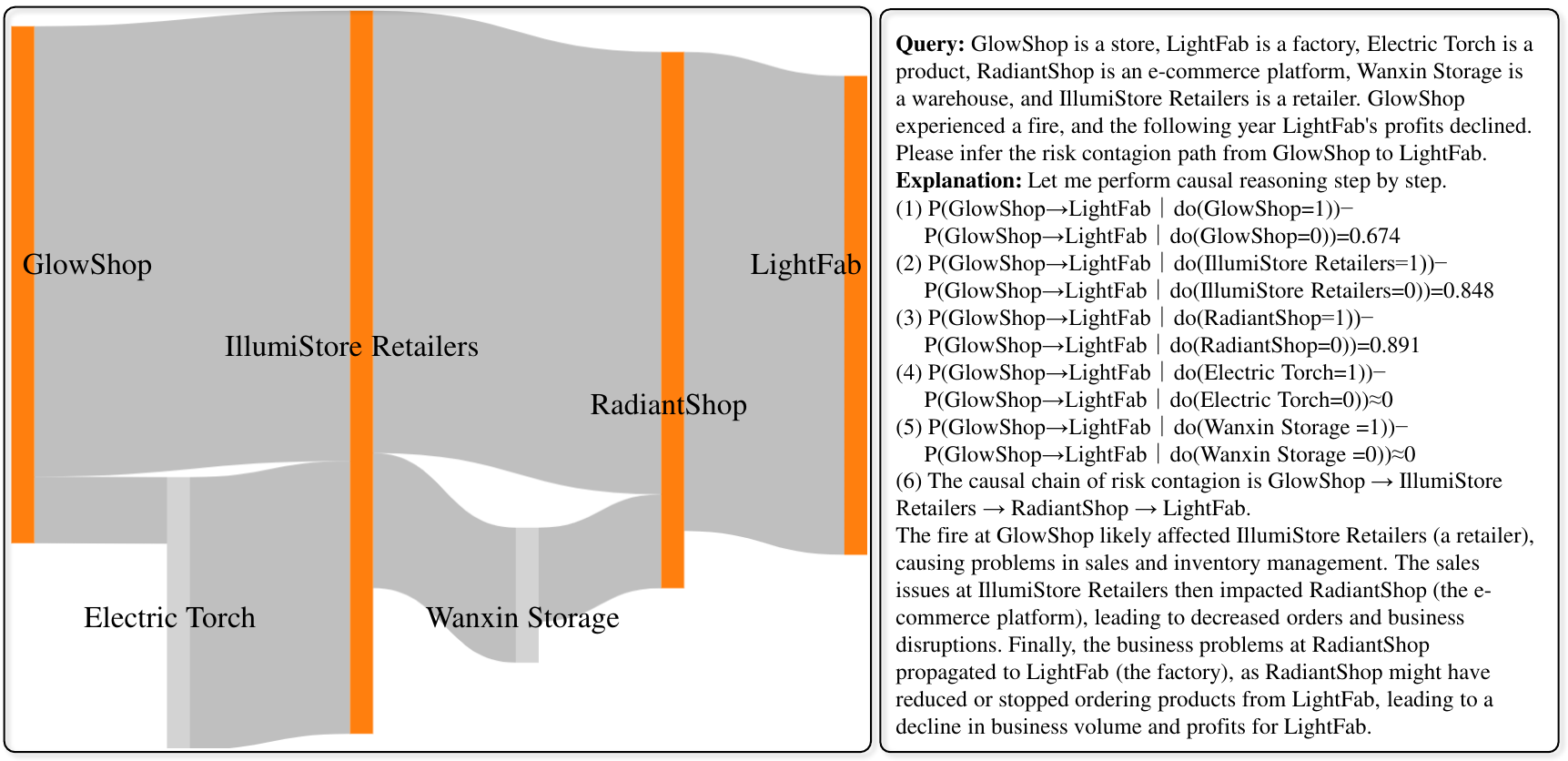}
    \caption{Visualization of Risk Intensity via a Sankey Diagram \& Causal Explanations: Case $1$}
    \label{fig:Sankey-1}
\end{figure*}

\begin{figure*}[htbp]
\centering 
    \includegraphics[width=0.8\textwidth]{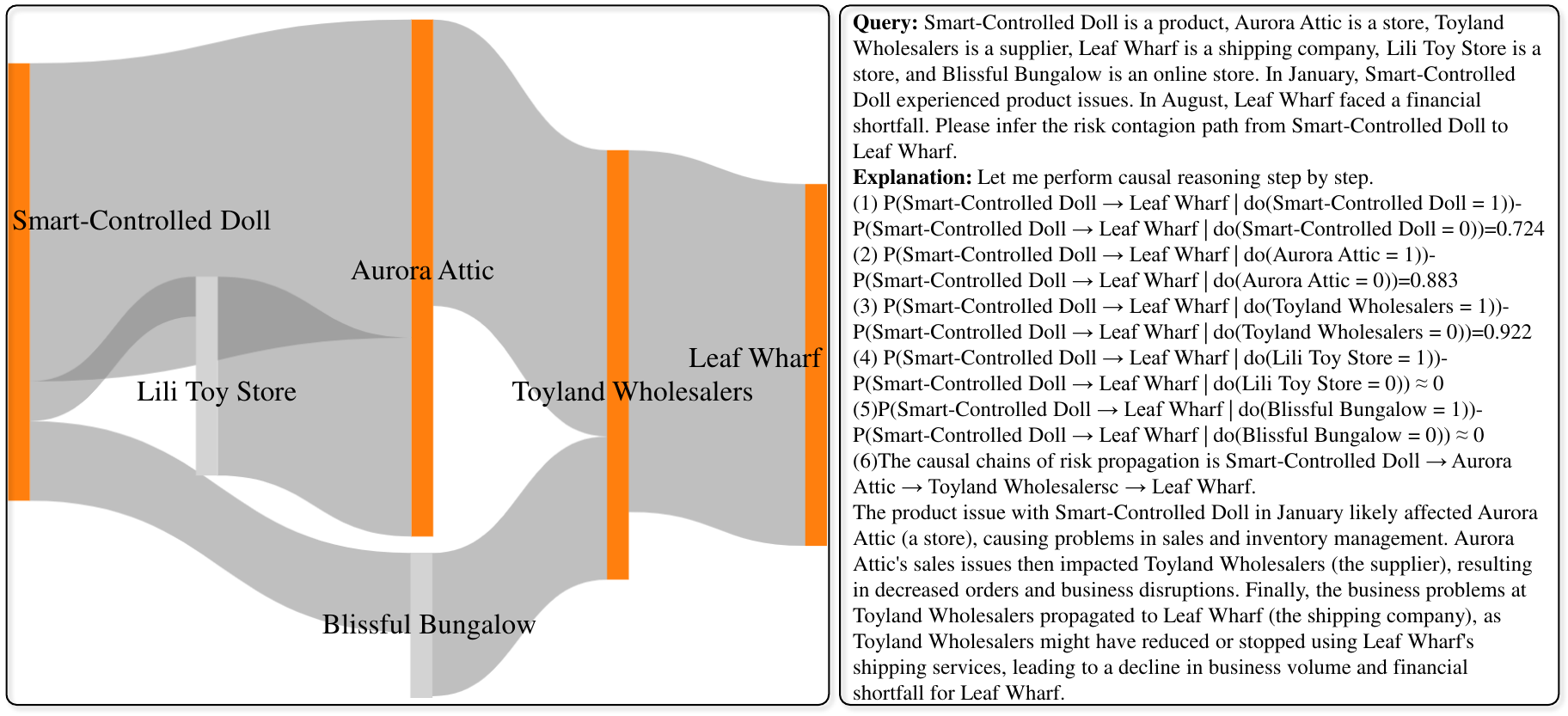}
    \caption{Visualization of Risk Intensity via a Sankey Diagram \& Causal Explanations: Case $2$}
    \label{fig:Sankey-2}
\end{figure*}

\begin{figure*}[htbp]
\centering 
    \includegraphics[width=0.8\textwidth]{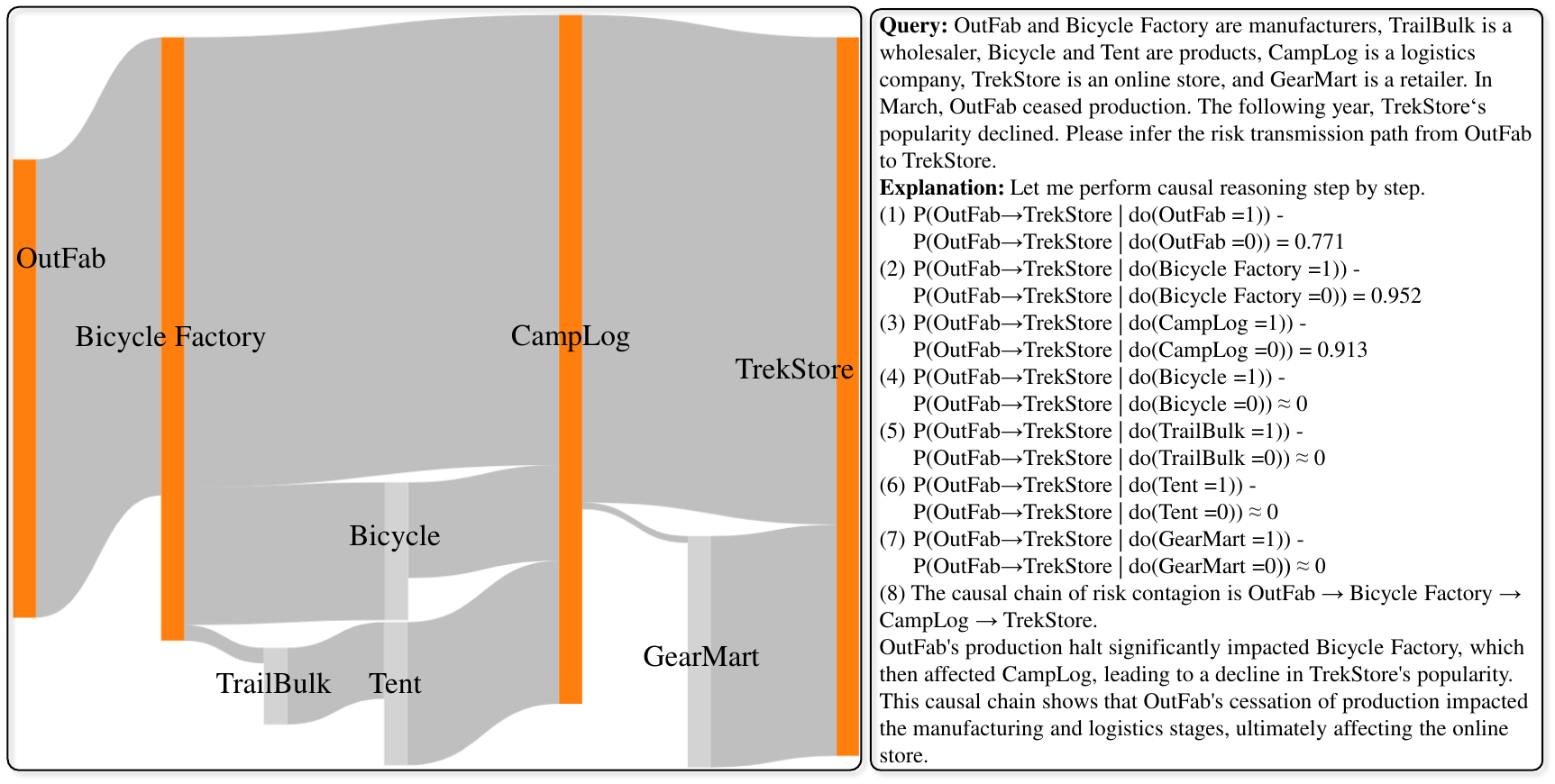}
    \caption{Visualization of Risk Intensity via a Sankey Diagram \& Causal Explanations: Case $3$}
    \label{fig:Sankey-3}
\end{figure*}

\begin{sloppypar}
As illustrated in Fig.~\ref{fig:Sankey-1}, our model provides detailed explanations when answering queries and visualizes the intensity and direction of financial risk contagion. The diagram highlights the path of risk contagion through orange nodes, specifically
GlowShop (store)$\rightarrow$ IllumiStore Retailers (retailer)$\rightarrow$ RadiantShop (e-commerce platform) $\rightarrow$ LightFab (factory), which is the actual risk contagion path in the SupplyChain-KG dataset. 
The thickness of the path indicates the intensity of the risk contagion.
For more cases, please refer to Figs.~\ref{fig:Sankey-2} and~\ref{fig:Sankey-3}.
\end{sloppypar}

\subsection{Ablation Study}
\subsubsection{Performance of Combining LLMs with KGs}
From Table~\ref{tab:answers}, we observe significant enhancements in ACC for both datasets through the LLMs+KGs approach, with increases of $11.1\%$ and $6.529\%$, respectively. 
In Table~\ref{tab:ood-answers}, the LLMs+KGs methodology showcases substantial improvements in managing OOD data over the use of LLMs alone. Specifically, it achieved ACC performance boosts of $13.779\%$ and $5.879\%$ for the FinDKG and SupplyChain-KG datasets, respectively, along with AUC performance enhancements of $12.033\%$ and $4.9\%$.
These results underscore the efficacy of integrating KGs in boosting prediction accuracy, particularly in contexts requiring nuanced domain knowledge. This enhancement is likely attributed to the structured knowledge from KGs, which supports our model in achieving a deeper understanding and more effective reasoning.

In Table~\ref{tab:answers}, the LLMs+KGs approaches demonstrate a significant performance enhancement over the GNNs methods, with ACC improvements of $12.5\%$ and $7.2\%$ for the FinDKG and SupplyChain-KG datasets, respectively, alongside AUC gains of $12.65\%$ and $9.7\%$. 
From Table~\ref{tab:ood-answers}, it is evident that the LLMs+KGs approaches markedly surpass the GNNs methods in addressing OOD data, showcasing ACC and AUC enhancements of $16.2\%$ and $16.3\%$ for the FinDKG dataset, and $5.15\%$ and $4.95\%$ for the SupplyChain-KG dataset, respectively. 
These significant improvements are primarily attributable to the reasoning and generalization capabilities of LLMs.

\subsubsection{Performance of Cross Attention Mechanisms}
In Table~\ref{tab:answers}, we observe that removing multi-head attention mechanisms results in decreased ACC and AUC metrics. Specifically, for the FinDKG dataset, the reductions in ACC and AUC are approximately $7.3\%$ and $8.6\%$, respectively; for the SupplyChain-KG dataset, the decreases are around $4.5\%$ and $2\%$, respectively.
These findings suggest that multi-head attention mechanisms significantly affect the integration of KGs with LLMs, thereby impacting model performance.

\subsubsection{Performance of Multi-scale Contrastive Loss}
In Table~\ref{tab:answers}, 
%
%
compared to $\text{RC\textsuperscript{2}R}^\ddagger$, RC\textsuperscript{2}R achieves approximately $6.6\%$, $6\%$, and $3.5\%$, $1.1\%$ improvements in ACC and AUC performance on the FinDKG and SupplyChain-KG datasets, respectively. 
This significant leap in performance can largely be attributed to the multi-scale contrastive learning function, which effectively aligns fine-grained information between text and graphs.

\subsubsection{Robustness Analysis of Causal Reasoning}
To evaluate the robustness of our model's causal reasoning against the risk of contagion, we perform two tests on FinDKG: the \textit{Random Confounder} and the \textit{Subset of Data} tests~\cite{kiciman2018tutorial}.
In the \textit{Random Confounder} test, we introduce additional confounding nodes into the factual graph to simulate potential external influences.
The \textit{Subset of Data} test involves the random removal of nodes from the factual graph, which helps us understand how our model copes with incomplete data sets.

\begin{figure}[htbp]
\centering 
    \includegraphics[width=0.5\textwidth]{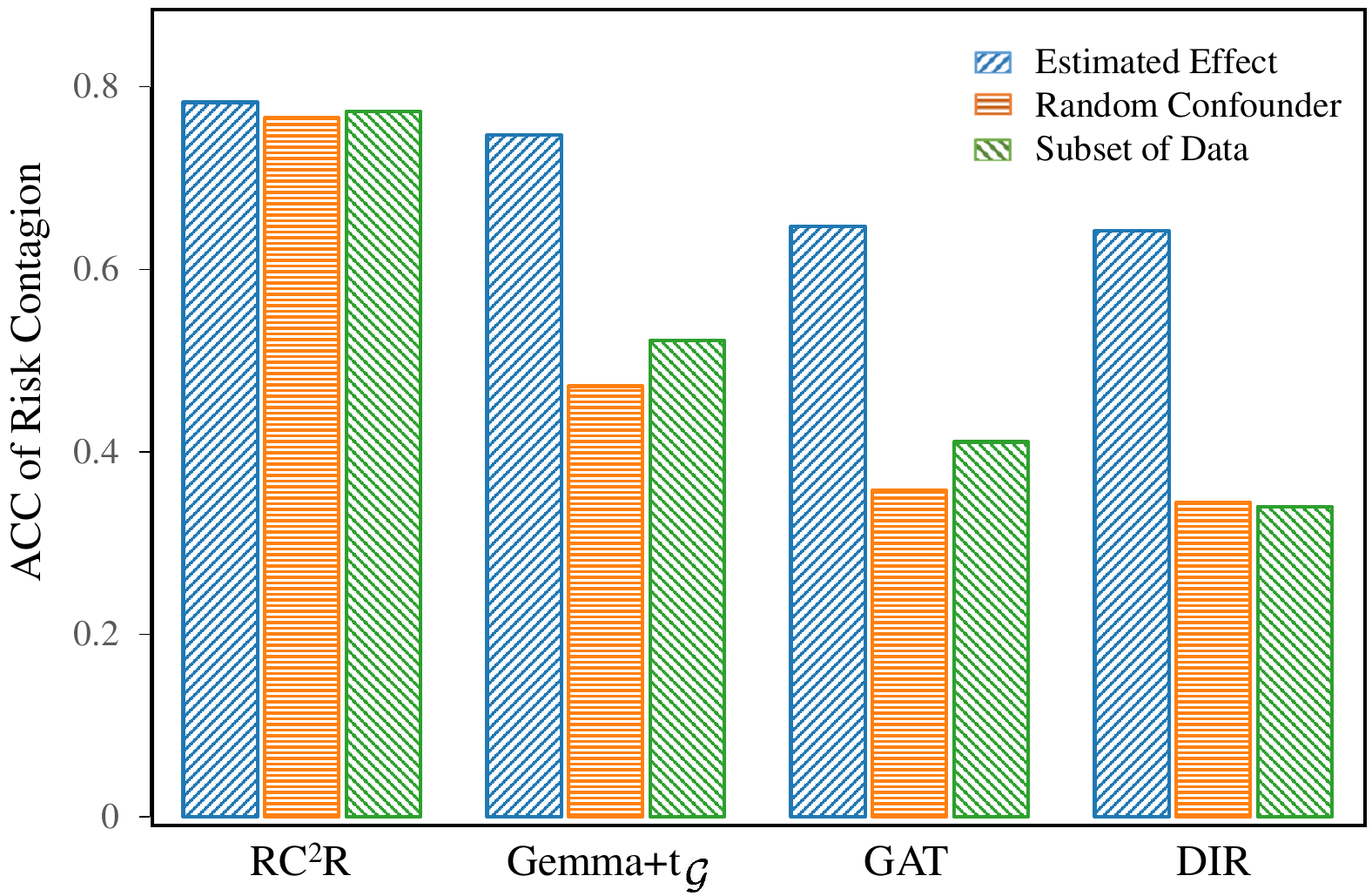}
    \caption{Robustness Test of Causal Reasoning on FinDKG}
    \label{fig:Quality-CG}
\end{figure}

As shown in Fig.~\ref{fig:Quality-CG}, our model maintains consistent performance in the \textit{Random Confounder} and \textit{Subset of Data} tests compared to the \textit{Estimated Effect} test, whereas the other three models exhibit significant declines under the same conditions. This underscores the robust causal reasoning of our model. Such robustness primarily stems from the diverse interventions we implemented and the explicit incorporation of influence estimation for fine-grained causal features in \(\mathcal{L}_{path}\). In contrast, Gemma+\(t_g\) and DIR lack these design mechanisms. Additionally, the attention mechanism employed by GAT is based on correlations that lack the robustness of causal inference.

\section{Conclusion}

To analyze the causal effects behind risk contagion,
this study proposes RC\textsuperscript{2}R, which combines the reasoning capabilities of LLMs with the factual and professional knowledge within KGs, aiming to infer the causal relationships in financial risk contagion.
At the data level, we rigorously follow the hierarchies of causation to develop causal instructions for fine-tuning our model and activating the causal reasoning capabilities of LLMs.
In terms of model architecture level, we develop a fusion module integrating textual and graph information through a multi-scale contrastive loss function and soft prompt with multi-head attention mechanisms. 
Furthermore, we establish a risk pathway inference module
to identify propagation routes, which are visualized via
Sankey diagrams.
%
In the near future, we intend to develop agents grounded in LLMs to autonomously conduct formal causal reasoning. We aim to apply this technology to analyze the causal factors behind financial risk contagion~\cite{algieri2017assessing,gong2022geopolitical}.

\bibliographystyle{IEEEtran}
\bibliography{ref.bib}

\vfill

\end{document}